\documentclass[sigconf,natbib=true,anonymous=false]{acmart}

\AtBeginDocument{%
  \providecommand\BibTeX{{%
    \normalfont B\kern-0.5em{\scshape i\kern-0.25em b}\kern-0.8em\TeX}}}

\setcopyright{acmcopyright}
\copyrightyear{2023}
\acmYear{2023}
\acmDOI{XXXXXXX.XXXXXXX}

\acmConference[The Web Conference '23]{The Web Conference}{April 30--May 04,
  2023}{Austin, Texas}
\acmISBN{978-1-4503-XXXX-X/18/06}

\usepackage{amssymb}
\usepackage{multirow} 
\begin{document}

\title{Adaptive Utilization of Cross-scenario Information for Multi-scenario Recommendation}



\author{Xiufeng Shu, Ruidong Han, Xiang Li, Wei Lin}
\affiliation{%
  \institution{Meituan}
  \city{Beijing}
  \country{China}}
\email[email1]{xf_shu95@163.com, hanruidong@meituan.com}

\renewcommand{\shortauthors}{Xiufeng Shu, et al.}

\begin{abstract}
Recommender system of the e-commerce platform usually serves multiple business scenarios. Multi-Scenario Recommendation (MSR) is an important topic that improves ranking performance by leveraging information from different scenarios. Recent methods for MSR mostly construct scenario shared or specific modules to model commonalities and differences among scenarios. However, when the amount of data among scenarios is skewed or data in some scenarios is extremely sparse, it is difficult to learn scenario-specific parameters well. Besides, simple sharing of information from other scenarios may result in negative transfer. In this paper, we propose a unified model named {\bfseries Cross-Scenario Information Interaction (CSII) } to serve all scenarios by a mixture of scenario-dominated experts. Specifically, we propose a novel method to select highly transferable features in data instances. Then, we propose an attention-based aggregator module, which can adaptively extract relative knowledge from cross-scenario. Experiments on the production dataset verify the superiority of our method. Online A/B test in Meituan Waimai APP also shows a significant performance gain, leading to an average improvement in GMV (Gross Merchandise Value) of 1.0\% for overall scenarios.
\end{abstract}

\ccsdesc[500]{Information systems~Recommender systems}

\keywords{Recommender System, Multi-Scenario Recommendation, Neural Networks }


\maketitle

\begin{figure*}
  \includegraphics[width=0.95\textwidth]{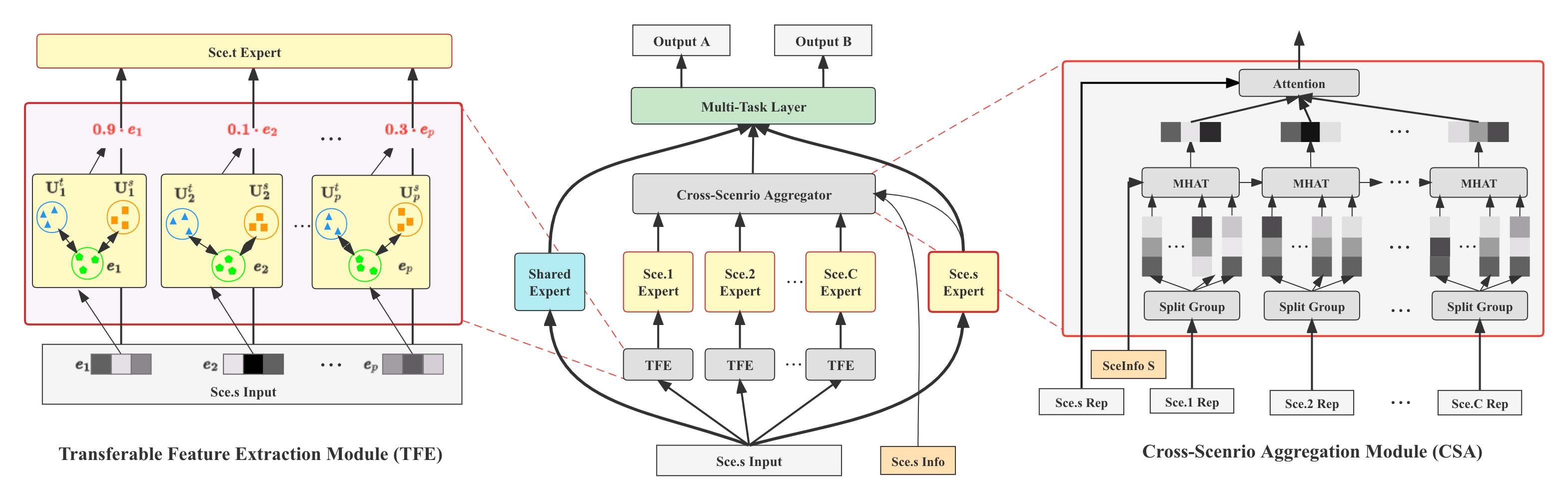}
  \setlength{\abovecaptionskip}{0.cm}
  \captionsetup{font={small}}
  \caption{The structure of Cross-Scenario Information Interaction (CSII), which consists of two key modules: Transferable Feature Extraction (TFE) and Cross-Scenario Aggregation (CSA).}
  \Description{Enjoying the baseball game from the third-base
  seats. Ichiro Suzuki preparing to bat.}
  \label{fig:arch}
\end{figure*}

\section{Introduction}
Recommender System (RS) of e-commerce platform usually serves multiple business scenarios\cite{resnick1997recommender,lu2015recommender}, such as Meituan, Amazon, etc. In Meituan Waimai App (the largest food delivery platform in China), a scenario refers to a certain channel such as Homepage, Food channel, Dessert channel, etc. These scenarios are separated by time period, category, and other business factors. RS usually needs to predict Click-Through Rate (CTR) and post-view Click-Through\&Conversion Rate (CTCVR) for multiple scenarios using Multi-Task Learning (MTL)\cite{zhao2019recommending,caruana1997multitask,ma2018modeling,ma2018entire}.   

In this work, we focus on Multi-Scenario Recommendation (MSR), which aims to enhance ranking ability by utilizing the information from different scenarios. Recently, significant efforts\cite{li2020improving,sheng2021one,shen2021sar} have been devoted to devising different MSR methods. Some studies like STAR\cite{sheng2021one} propose a shared module and a scenario-specific module to extract common knowledge and scenario-specific knowledge respectively. Moreover, they can avoid the problem that the minor scenario may be dominated by the 
major scenario. Besides, the information utilization of cross-scenario in MSR is the reason for the performance improvement of the model. However, previous works suffer from two crucial problems:

\textbf{(1) Insufficient exploitation of data instances from different scenarios}: 
In most methods, the scenario-specific module can only use the data instances of one scenario during training. However, when the amount of data among scenarios is skewed or the data in some scenarios is extremely sparse, the scenario-specific module is usually underfitted\cite{kaplan2020scaling}. Augmenting data from other scenarios to help convergence is an intuitive method, but it may cause negative transfer\cite{pan2009survey,yao2010boosting} due to the difference in features and data distribution among scenarios. Therefore, how to choose shareable features and highly relevant instances for a certain scenario is an important problem.

\textbf{(2) Selective aggregation of cross-scenario representation problem}: 
Although existing methods\cite{li2020improving,sheng2021one,shen2021sar} can exploit cross-scenario knowledge by utilizing the scenario-shared module, they rarely use the knowledge from the other scenario-specific module that result in the waste of information. If users and items are fully overlapped in various scenarios, the instances can be sent to the module of other scenarios to extract potential relevant knowledge and use it to improve model accuracy. In addition, different instances should pay various importances to the knowledge among scenarios, and most methods ignore the selection of information. It is necessary to dynamically aggregate the information according to the input scenario.

 To solve the above issues, we propose a {\bfseries Cross-Scenario Information Interaction model (CSII)} in this paper. We construct scenario-dominated experts and a shared expert in our model to leverage multi-scenario information. Specifically, we propose a Transferable Feature Extraction module (TFE) to select highly transferable features among scenarios and then feed the results of TFE into all the other scenario-dominated experts. Finally, we propose a Cross-Scenario Aggregation module (CSA), which uses a multi-level attention mechanism to aggregate the information from different experts. The main contributions of this work are summarized as follows:
\begin{itemize}
\item We propose a Transferable Feature Extraction module to effectively avoid the negative transfer issue in cross-scenario data utilization.
\item We propose a Cross-Scenario Aggregation Module to model the relationship among scenarios, and extract scenario-related information in the representations.
\item We conduct extensive experiments on Meituan real-world large-scale recommendation datasets. Both offline and online experiments demonstrate the superiority of our proposed CSII. Currently, CSII has been successfully deployed in Meituan to serve all scenarios.
\end{itemize}

\section{ CROSS-SCENARIO INFORMATION INTERACTION MODEL }
In this section, we introduce the overall architecture of our proposed model. CSII is a general and flexible architecture that can treat different scenarios in a unified framework. As shown in Figure.\ref{fig:arch}, CSII consists of two key modules. A module named Transferable Feature Extraction (TFE) for measuring the transferability of each feature,  and a module named  Cross-Scenario Aggregation (CSA) for aggregating knowledge from the mixture of scenario-dominated experts. We will describe the two modules in detail.

\subsection{Transferable Feature Extraction Module}
We assume that there are C scenarios. Each sample from a specific scenario $s$ has $P$ different categorical features, denoted as $\pmb{x}=[x_1, \dots, x_p]$. $\pmb{e}_{p} \in \mathbb{R}^d$ is the corresponding representation of $x^s$ using embedding layer $E(\cdot)$(we also assume that the same feature has the same embedding representation among all the scenarios). For each scenario pair $(s,t)$ \,where $1 \leq s,t \leq C $, we hope to learn a set of transformation functions $\psi_p^{st} :\mathbb{R}^d \rightarrow\mathbb{R}^d$ which can improve the transferability of the $p$-th feature from scenario $s$ to scenario $t$.

Inspired by\cite{wang2019transferable,swietojanski2016learning}, we learn the transformation functions by measuring the discrepancy between two selected scenarios using the more flexible and parameter-based method named TFE. For each feature, we define a set of learnable parameters
$\{ \mathbf{U}_p^{t}\}_{1 \leq t \leq C}$ where $\mathbf{U}_p^{t} \in \mathbb{R}^d$. Then, we measure the relevance of the $p$-th feature to scenario $t$ by calculating the distance between the feature's embedding $\pmb{e}_{p}$ and the vector $\mathbf{U}_p^t$:
\begin{equation}
D(\pmb{e}_p,t) \triangleq ||\pmb{e}_p - \mathbf{U}_p^{t}||_2^2
\end{equation}
where we use euclidean distance as the distance function and some other metrics like cosine similarity can also be considered. 

Before data instance from scenario $s$ is fed into the expert with respect to the scenario $t$, we compute both $D(\pmb{e}_p,s )$ and $D(\pmb{e}_p,t )$ and define a transferability score to adjust the weight of feature in data instance:
\begin{equation}
w_p^{st}(\pmb{e}_p) = \mathrm{exp}(-\left |  D(\pmb{e}_p,s ) - D(\pmb{e}_p,t)\right | ) 
\end{equation}
As shown in the above formula, if the two distances are different,  the $w_p^{st}(\pmb{e}_p)$ is a small value, which means that the expression of this feature in the two scenarios is different. In particular, if $s=t$, the $w_p^{st}(\pmb{e}_p)$ is 1. Furthermore, the final transformation $\psi(\cdot)$ is given by: 
\begin{equation}
\psi_p^{st}(\pmb{e}_p) = 2 \cdot \sigma (\alpha  \cdot w_p^{st}(\pmb{e}_p) + \beta
) \cdot \pmb{e}_p
\end{equation}
where $ \alpha$ and $\beta$ are feature-scenario-aware parameters that are used to adjust the feature weight in different scenarios. $\sigma(\cdot)$ is sigmoid function. Using TFE, all features in the data instance of the scenario $s$ can get its transferable part:
\begin{equation}
\begin{aligned}
\pmb{v}^{st}  =f_t(\mathrm{Concat} (\psi_1^{st}(\pmb{e}_1), \psi_2^{st}(\pmb{e}_2 ),...,
\psi_P^{st}(\pmb{e}_P))
) 
\end{aligned}
\end{equation} 
where $f_t$ is an MLP-based scenario expert, and we will show the scenario-dominated property of this expert structure in section 2.3.

For each sample $\pmb{x}$ from scenario $s$, we emphasize that we use TFE for all the scenarios rather than one specific scenario at the same time and we can get a set of representations  $\{\pmb{v}^{st}\}_{1\leq  t \leq C}$ to help each scenario expert to achieve better performance.
 
\subsection{Cross-Scenario Aggregation Module}
As mentioned above, we use the TFE module to improve the transferability in feature level for each scenario pair and get more transferable results. Then, we take two aggregators in different levels to process these results and obtain the aggregated information, which is used as input for the MLP-based classifier.
\paragraph{\textbf{Intra-Scenario Transferability Aggregator  (Intra-Agg).}} It is worth noting that not all of the information in $\pmb{v}^{st}$ is useful for a specific scenario $t$. In order to aggregate these representations and mitigate the potential risk of negative transfer, we introduce a transformer-based\cite{vaswani2017attention} aggregation method to extract more relative knowledge. We define $\pmb{K}=\pmb{V}=\pmb{v}^{st}$ and the only difference is that the query vector consists of the factors about the current input scenario. For example, in our work, we use channel indicator, mealtime, location, and category as query feature, which is defined as 
$\pmb{Q}=\mathsf{Concat}(\pmb{e}_{i_1},\pmb{e}_{i_2},...,\pmb{e}_{i_q})$, where ($i_1,i_2,...,i_q$) are the index from the subset of the features. Then, multi-head self-attention for each scenario $t$ can be formulated as:

\begin{align}
\pmb{h}_{t}=\mathsf{Concat}(\mathsf{head}_{t}^1,\mathsf{head}_{t}^2,...,\mathsf{head}_{t}^h)\pmb{W}_T^O \\
\mathsf{head}_{t}^i=\mathsf{Attention}(\pmb{Q}\pmb{W}_{T}^{Q_i},\pmb{K}\pmb{W}_{t}^{K_i},\pmb{V}\pmb{W}_{t}^{V_i}) \\
\mathsf{Attention}(\pmb{Q},\pmb{K},\pmb{V}) =\mathsf{softmax}(\frac{\pmb{Q}\pmb{K}^T}{\sqrt{d_K}})\pmb{V}
\end{align}

where $\pmb{W}_{t}^{K_i},\pmb{W}_{t}^{V_i} \in \mathbb{R}^{dP\times d_k} $, $\pmb{W}_{t}^{Q_i} \in \mathbb{R}^{dq\times d_k}$, $\pmb{W}_{t}^{O} \in \mathbb{R}^{hd_k\times dP} $ are parameter matrice and $d_k = dP/h$.
\paragraph{\textbf{Inter-Scenario Transferability Aggregator  (Inter-Agg).}} When we perform the intra-scenario aggregator above to get a scenario feature matrix, denoted as  $\tilde{\pmb{H}} = (\pmb{h}_{1}, \pmb{h}_{2}, ...,\pmb{h}_{C})$ from sample $\pmb{x}^s$, an inter-scenario aggregator using target attention can combine information from different scenarios to prevent negative transfer:
\begin{equation}
\pmb{u}_{agg} = \mathsf{Attention}(\tilde{\pmb{H}}, \pmb{h}_{s},\tilde{\pmb{H}})
\end{equation} 
where we use  $\pmb{h}_s$ as $\pmb{K}$ in the attention function, because knowledge from the current scenario of data instance is more important.
\subsection{Scenario-Dominated Paradigm \& Prediction}
Different from the scenario-specific paradigm, the scenario-dominated paradigm means samples from any scenario can be fed into each expert. Meanwhile, each expert is dominated by a concrete scenario. It plays an important role in our approach to taking experts to learn the discrepancy among scenarios. To this purpose, we use a scenario residual layer to increase the importance of the self-scenario information from the data instance in the learning process. We also use a shared expert to improve performance. In this way, we concatenate these two parts into high-dimension vector $\pmb{z}$ and as the input of multi-task layers:
\begin{equation}
\pmb{z} =  \mathsf{Concat}(\alpha_{s} \cdot \pmb{u}_{agg} + \pmb{h}_{s}, \pmb{h}_{share})
\end{equation}
where $\pmb{h}_{share}$ is the output from the shared expert and $\alpha_S$ is the weight to adjust strength about other scenarios expressions. 

Finally, We can use arbitrary multi-task methods to make predictions. For simplification, we directly feed $\pmb{z}$ into a shared MLP $\phi_k$ to generate the prediction for each task $k$:
\begin{equation}
\hat{y}_{k}^i=\mathrm{sigmoid}(\phi_k(\mathsf{\pmb{z}}))  
\end{equation}

We utilize the standard cross-entropy loss function to optimize each task, including CTR and CTCVR.
\begin{table*}[]
\small
\setlength\tabcolsep{3pt}
\captionsetup{font={small}}
\caption{Performance comparison in Meituan production dataset. The best results are in boldface and second best underlined. All experiments are repeated 3 times and averaged results are reported. The evaluation metric A-TR refers to the AUC of CTR, and A-VR refers to the AUC of CTCVR.}
\centering
\label{overall_pe}
\vspace{-1.0em}

\scalebox{1.0}{
\setlength{\tabcolsep}{1.7mm}{
\begin{tabular}{llcccccccccccccc}
\toprule
 \multicolumn{2}{c}{\multirow{2}{*}{Method}} & \multicolumn{2}{c}{Sce.1} & \multicolumn{2}{c}{Sce.2} & \multicolumn{2}{c}{Sce.3} & \multicolumn{2}{c}{Sce.4} & \multicolumn{2}{c}{Sce.5} & \multicolumn{2}{c}{Sce.6} & \multicolumn{2}{c}{Sce.7} \\ 
\cmidrule(lr){3-4} \cmidrule(lr){5-6} \cmidrule(lr){7-8} \cmidrule(lr){9-10} \cmidrule(lr){11-12} \cmidrule(lr){13-14} \cmidrule(lr){15-16}
\multicolumn{2}{c}{~} & \multicolumn{1}{c}{\begin{tabular}[c]{@{}c@{}}A-TR\end{tabular}} & \multicolumn{1}{c}{\begin{tabular}[c]{@{}c@{}}A-VR\end{tabular}} & \multicolumn{1}{c}{\begin{tabular}[c]{@{}c@{}}A-TR\end{tabular}} & \multicolumn{1}{c}{\begin{tabular}[c]{@{}c@{}}A-VR\end{tabular}} & \multicolumn{1}{c}{\begin{tabular}[c]{@{}c@{}}A-TR\end{tabular}} & \multicolumn{1}{c}{\begin{tabular}[c]{@{}c@{}}A-VR\end{tabular}} & \multicolumn{1}{c}{\begin{tabular}[c]{@{}c@{}}A-TR\end{tabular}} & \multicolumn{1}{c}{\begin{tabular}[c]{@{}c@{}}A-VR\end{tabular}} & \multicolumn{1}{c}{\begin{tabular}[c]{@{}c@{}}A-TR\end{tabular}} & \multicolumn{1}{c}{\begin{tabular}[c]{@{}c@{}}A-VR\end{tabular}} & \multicolumn{1}{c}{\begin{tabular}[c]{@{}c@{}}A-TR\end{tabular}} & \multicolumn{1}{c}{\begin{tabular}[c]{@{}c@{}}A-VR\end{tabular}} & \multicolumn{1}{c}{\begin{tabular}[c]{@{}c@{}}A-TR\end{tabular}} & \multicolumn{1}{c}{\begin{tabular}[c]{@{}c@{}}A-VR\end{tabular}} \\ \hline
\multicolumn{2}{l}{MTL Base \cite{caruana1997multitask}} & 0.6825 & 0.7053 & 0.6810 & 0.7317 & 0.7208 & 0.7169 & 0.6851 & 0.6817 & 0.6413 & 0.6249 & 0.6567 & 0.6500 & 0.6746 & 0.6650 \\
\multicolumn{2}{l}{MMoE\cite{ma2018modeling}} & 0.6846 & 0.7078 & 0.6836 & 0.7344 & 0.7232 & 0.7187 & 0.6890 & 0.6845 & 0.6446 & 0.6280 & 0.6600 & \underline{0.6514} & 0.6781 & 0.6688 \\
\multicolumn{2}{l}{HMoE\cite{li2020improving}} & 0.6830 & 0.7081 & 0.6833 & 0.7333 & \underline{0.7233} & 0.7178 & 0.6889 & 0.6847 & 0.6430 & 0.6260 & 0.6590 & 0.6505 & 0.6778 & \underline{0.6692} \\
\multicolumn{2}{l}{PLE\cite{tang2020progressive}} & \underline{0.6855} &  \underline{0.7085} & \textbf{0.6851} & \underline{0.7348} & 0.7231 & \underline{0.7188} & \underline{0.6892} & \underline{0.6851} & \underline{0.6449} & \underline{0.6287} & \underline{0.6610} & 0.6511 & \underline{0.6784} & 0.6607 \\
\multicolumn{2}{l}{STAR\cite{sheng2021one}} & 0.6833 & 0.7056 & 0.6821 & 0.7311 & 0.7223 & 0.7175 & 0.6862 & 0.6811 & 0.6421 & 0.6244 & 0.6573 & 0.6482 & 0.6764 & 0.6676 \\
\multicolumn{2}{l}{\textbf{CSII}} & \textbf{0.6858} & \textbf{0.7104} & \underline{0.6848} & \textbf{0.7359} & \textbf{0.7247} & \textbf{0.7199} & \textbf{0.6910} & \textbf{0.6875} & \textbf{0.6457} & \textbf{0.6319} & \textbf{0.6613} & \textbf{0.6554} & \textbf{0.6803} & \textbf{0.6729} \\
\bottomrule
\end{tabular}
    }
}
\end{table*}

\section{EXPERIMENTS}
\begin{table}[!h]
\captionsetup{font={small}}
\caption{Statistics on the Meituan dataset.}
\vspace{-1.0em}
\centering
\label{expess_rate}

\scalebox{0.8}{

\begin{tabular}{lccccccc}
\toprule
 & Sce.1 & Sce.2 & Sce.3 & Sce.4 & Sce.5 & Sce.6 & Sce.7 \\
\hline
sample percentage & 71.73\% & 18.60\% & 5.55\% & 1.47\% & 1.42\% & 0.94\% & 0.29\% \\
average CTR & 6.18\% & 6.02\% & 9.83\% & 8.43\% & 7.05\% & 7.13\% & 7.18\% \\
average CTCVR & 0.89\% & 0.63\% & 1.27\% & 0.81\% & 0.92\% & 0.64\% & 0.80\% \\
\bottomrule
\end{tabular}

 }
\end{table}
In this section, we evaluate our proposed CSII against a series of state-of-the-art baselines. Extensive experiments on real-world large-scale datasets demonstrate the effectiveness of our model, which is further confirmed by the online A/B test across multiple business metrics.
\subsection{Experimental Settings}
{\textbf{Dataset.} We collect a production dataset from Meituan Waimai APP to perform the offline evaluation. The dataset is collected from the log of the recommender system from  July. 01 to July. 30 2021, which has billion of samples per day and consists of 7 business scenarios. As shown in Table \ref{expess_rate}, the proportion of data in different business scenarios is seriously unbalanced, the major scenario occupies 70\% of the exposure, and minor scenarios account for less than 1\% of the exposure.
 
\textbf{Baselines.}
To verify the effectiveness of the proposed approach, we compare CSII with the following models:
\begin{itemize}
\item {\textbf{MTL Base}\cite{caruana1997multitask}.}We use a simple classical multi-task model, in which all tasks share the embedding layer at the bottom and each task has a specific network at the top.
\item {\textbf{MMoE}\cite{ma2018modeling}.} This method designs multi-gate mixture experts to implicitly model the relationship between tasks. Besides, We use scenario-related information as the input of gate-network to improve model performance.
\item {\textbf{HMoE}\cite{li2020improving}.} HMoE extends MMoE to scenario-aware experts with the gradient-cutting trick for encoding scenario correlation explicitly.
\item {\textbf{PLE}\cite{tang2020progressive}.} PLE extends MMoE with separates experts into task-specific groups. We adopt it by separating experts into scenario-specific groups.
\item {\textbf{STAR}\cite{sheng2021one}.} STAR proposes a star topology that consists of a shared center and scenario-specific parameters. We implement this star topology for each task.
\end{itemize}
\textbf{Metrics and Implementation.}
In this work, we use Adam\cite{kingma2014adam} as the optimizer with a learning rate of 0.001 and mix samples from different scenarios to train those baselines. All models use DIN\cite{zhou2018deep} modules to process user behavior sequences and have the same network depth and width. In MMoE and HMoE, the number of experts is the same as the number of scenarios. We use the ROC curve (AUC) as the metric to evaluate the performance both of CTR and CTCVR in each scenario.

\subsection{Overall Experimental Results} 
As shown in Table.\ref{overall_pe}, 
for scenarios such as Sce.5 and Sce.6 where data is severely sparse, STAR is not as effective as MTL Base, mainly because of the poor performance of the scenario-specific modules. MMoE which we implemented allows each scenario to share a set of experts, and it can be observed that the performance of each scenario is better than STAR and MTL Base. Compared to PLE, although we only have one expert per scenario, achieves 0.004 better performance with 30\% parameter reduction. Meanwhile, it is notable that in commercial RS with billions of impressions, 0.001 absolute AUC gain is significant in our baseline with thousands of features and complex user behavior modules.

\begin{table}[!h]
\vspace{-1.0em}
\captionsetup{font={small}}
\caption{Ablation study of TFE.}
\vspace{-1.0em}
\centering
\label{tfe}

\scalebox{0.8}{
\setlength{\tabcolsep}{2.6mm}{
\begin{tabular}{lcccc}
\toprule
 & PLE & PLE w/ TFE & CSII w/o TFE & CSII \\
\hline
Overall AUC CTR & 0.6810 & 0.6813 & 0.6814 & 0.6819 \\
Overall AUC CTCVR & 0.6840 & 0.6850 & 0.6868 & 0.6878 \\

\bottomrule
\end{tabular}
}
 }
 \vspace{-1.0em}
\end{table}

\begin{table}[!h]
\captionsetup{font={small}}
\caption{Ablation study of CSA. 
}
\vspace{-1.0em}
\centering
\label{csa_result}

\scalebox{0.75}{

\begin{tabular}{lccccc}
\toprule
 & CSII$\dag$ &  CSII\dag w/ Intra-Agg&  CSII\dag w/ Inter-Agg & CSII\\
\hline
Overall AUC CTR & 0.6813 & 0.6815 & 0.6818 & 0.6819\\
Overall AUC CTCVR & 0.6847 & 0.6856 & 0.6872 & 0.6878\\

\bottomrule
\end{tabular}
 }
\end{table}

\subsection{Ablation Study} 

\subsubsection{Transferable Feature Extraction Module} We verify the efficiency of the TFE module in CSII. In addition, we apply it to a PLE model which is similar to our framework. The result is reported in Table.\ref{tfe}. Both PLE and our model achieve performance gains after applying TFE module. It is worth noting that TFE module only adds 1000 times fewer parameters than the overall model parameters.

\subsubsection{Cross-Scenario Aggregation Module}  We investigate the impact of Intra-Agg and InterAgg in the Cross-Scenario Aggregation Module. The base model is CSII\dag, which uses the mean pooling operator to replace the CSA module. After that, we build three models for comparison: 1) Base model with Intra-Agg. 2) Base model with Inter-Agg  3) Base model with both Inter-Agg and Intra-Agg (CSII). The experimental results in Table.\ref{csa_result} confirm that both  Intra-Agg and Inter-Agg are effective, and the effects can be cumulative. Additionally, we visualize the statistics of the score matrix output by attention in Inter-Agg. As can be seen from Figure.\ref{att}, the relationship among scenarios represented by attention is close to the real situation. For example, Sce.3 and Sce.7 are two similar channels (dessert and drink), and the above attention score takes a large value between them.

\begin{figure}[!t]
\centering
	\includegraphics[width=0.8\linewidth]{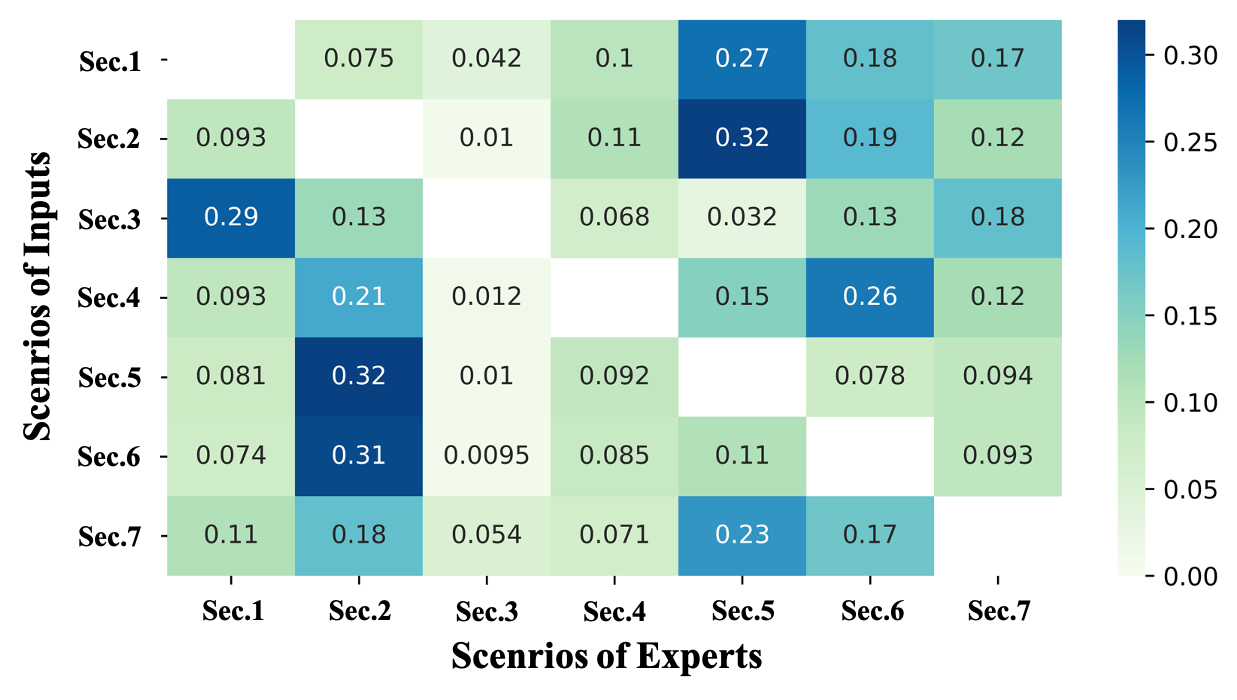}
    \setlength{\abovecaptionskip}{0.cm}
    \captionsetup{font={small}}
	\caption{The statistics of the score matrix  by attention in Inter-agg.}
    \label{att}
    
\end{figure}

\begin{table}[!h]
\vspace{-0.5em}
\captionsetup{font={small}}
\caption{The result of online A/B test in each scenario.}
\vspace{-1.0em}
\centering
\label{abtestres}
\scalebox{0.8}{
\begin{tabular}{lccccccc}
\toprule
 & Sce.1 & Sce.2 & Sce.3 & Sce.4 & Sce.5 & Sce.6 & Sce.7 \\
\hline
CTCVR & +1.02\% & +0.78\% & +0.40\% & +0.86\% & +1.41\% & +1.61\% & +1.57\% \\
GMV & +0.97\% & +0.84\% & +0.77\% & +0.99\% & +1.06\% & +1.28\% & +1.61\% \\
\bottomrule
\vspace{-1.0em}
\end{tabular}
 }
\end{table}

\subsection{Online A/B Test} 
We deploy CSII to the Meituan Waimai recommender system and conduct a two-weeks online A/B test on 7 important scenarios. Due to the large consumption of PLE online, the previous production model is deployed by MMoE. As shown in Table.\ref{abtestres}, compared to the previous one, the CSII has improved CTCVR by 1.02\% and GMV by 0.97\% in major scenarios. Especially for scenarios with sparse data, such as Sec.6-7, CSII has achieved an average increase of GMV of 1.47\%,  confirming that our method can well adapt to the problem of data skew. Considering the massive size of our system, such consistent online improvements are significant.

\section{CONCLUSION}
In this paper, we propose a Cross-Scenario Information Interaction model (CSII) to serve all scenarios by building scenario-dominated experts and a shared expert. Specifically, we first design a novel module to select highly transferable features in data instances to avoid negative transfer. Then, we propose an attention module to model scenario relationships, which can selectively aggregate knowledge among other scenarios. The experimental results from production data validate the superiority of the proposed CSII model. Since late 2021, The CSII model has been deployed as the ranking model in the Meituan Waimai APP recommender system and served 7 business scenarios, obtaining an increase of about 1.0\% in overall GMV.

\bibliographystyle{ACM-Reference-Format}
\bibliography{sample-base}

\appendix

\end{document}